\documentstyle[12pt]{article}
\begin{document}
\vspace*{-1in}
\newcommand {\lam}{\lambda'_{123}}
\newcommand {\phot} {\tilde{\gamma}}
\newcommand {\sel} {\tilde{e}_L}
\newcommand {\bsq} {\tilde{b}_R}
\renewcommand{\thefootnote}{\fnsymbol{footnote}}
\begin{flushright}
CERN-TH/95-320\\
CUPP-95/7\\
hep-ph/9512277 \\
\end{flushright}
\vskip 53pt
\begin{center}
{\Large{\bf \boldmath Searching $R$-parity-violating supersymmetry
                      in semileptonic $B$-decays}}
\vskip 30pt
{\bf Gautam Bhattacharyya\footnote{gautam@cernvm.cern.ch}}
\vskip 10pt
{\it Theory Division, CERN, \\ CH-1211  Geneva 23,  Switzerland}
\vskip 10pt
and
\vskip 10pt
{\bf Amitava Raychaudhuri\footnote{amitava@cubmb.ernet.in}}
\vskip 10pt
{\it Department of Pure Physics, University of Calcutta, \\
92 Acharya Prafulla Chandra Road, Calcutta 700 009, India.}
\vskip 70pt

{\bf ABSTRACT}
\end{center}

\begin{quotation}
If $R$-parity is broken and the photino, although
unstable, does not decay within the detector, then in new
semileptonic
$B$-decay modes a light ($\sim$ 2--3 GeV)  photino can be produced
carrying missing energy.  However, the photino, being massive,
arranges a different kinematical configuration for the visible decay
products as compared to a  standard semileptonic event where the
neutrino carries the missing energy.  We study the above kinematic
distributions in an attempt to explore the above scenario.

\end{quotation}

\vskip 100pt

\begin{flushleft}
CERN-TH/95-320\\
November 1995\\
\end{flushleft}

\setcounter{footnote}{0}
\renewcommand{\thefootnote}{\arabic{footnote}}
\vfill
\clearpage
\setcounter{page}{1}
\pagestyle{plain}

In the Minimal Supersymmetric Standard Model (MSSM), one of
the four neutralinos is the most favourable candidate to constitute
the lightest supersymmetric particle (LSP), which is assumed to be
stable. It is believed to weigh no less than $\sim$ 20 GeV
\cite{pdg} when
one compounds the non-observation of a gluino up to $\sim$ 120
GeV at the hadron colliders with the theoretical assumption of the
GUT-relation between the gaugino masses. But if
one relaxes the above assumption and if the LSP is dominantly
a gaugino (say, the photino), the above bound evaporates.
At low energy $e^+e^-$ machines light photinos can, in principle,
be pair produced in a $t$-channel selectron-exchanged process.
The anomalous single photon (ASP) experiment \cite{asp} was
designed to search for such a mode at the $e^+e^-$ storage ring
PEP, operating at $\sqrt{s} = 29$ GeV, at SLAC. A hierarchy
$m_{\phot} \ll \sqrt{s} \ll m_{\tilde{e}}$ was assumed in the
calculation of the radiative photino pair-production cross section.
The presence of single photons, with transverse momentum that cannot
be balanced by particles lost down the beam pipe, were demanded
to establish the ASP events. From the non-observation of such events
it was concluded that selectrons should be heavier than $\sim$ 60
GeV\footnote{ See Fig. 5 of
ref. \cite{asp} for the plots of
$\sigma(e^+e^-\rightarrow \gamma\phot\phot)$ as a function of
$m_{\tilde{e}}$ for various $m_{\phot}$ values.}.
On the other hand, cosmology demands that the cross section of
photino annihilation into electrons should not be too small,
requiring either the photino not to be too light or the selectron not
to be too heavy, to avoid the unwanted abundance of photinos in the
Universe. In this {\it tug of war} if we consider a photino
in the ball-park of $\sim$ 2--3 GeV
it is perhaps worth opening one more channel for it
to decay into standard particles by relaxing the assumption of
the so called ``$R$-parity conservation'' \cite{rpar}\footnote
{For an overview of various limits on the LSP, from accelerator
searches, dark matter searches and those of cosmological and
astrophysical origin, the reader is referred to {\it Supersymmetric
Particle Searches}, H. Haber in ref. \cite{pdg}. Of course, many of
these limits will not apply when a light photino is considered in
conjunction with $R$-parity violation.}.
In fact, such a light photino
should better decay within 1 s (which corresponds to a width
greater than $6.6\times 10^{-22}$ MeV) so that it does not live up
to the
nucleosynthesis era since injecting extra energy at this phase might
turn out to be unacceptable.  In this Letter we attempt to test the
existence of such a light photino in the context of semileptonic
$B$-decays \cite{cleolep}.
If $R$-parity is broken {\it carefully} so
that the photino remains stable within the detector, providing a new
funnel for carrying missing energy, it  can fake the Standard Model
(SM) neutrino in a semileptonic event.  First, we
summarize very briefly the basic features of explicit $R$-parity
violation (${\not \! \!{R}}$) and then discuss the strategy of
uncovering the above scenario.

\vskip 10pt

Normally searches for supersymmetric particles are carried out under
the assumption that a discrete quantum number $R$,
known as $R$-parity,
and defined as $R = (-1)^{(3B+L+2S)}$ where $B \equiv$ baryon number,
$L \equiv$ lepton number, and $S \equiv$ spin, is conserved. For all
ordinary particles $R=1$ and for all superparticles $R=-1$. However,
requiring the theory to be supersymmetric, renormalizable,
gauge-invariant and minimal in terms of field content does not
enforce
$R$-parity conservation.  Substituting in the Yukawa interaction the
$SU(2)$-doublet lepton superfield  in place of the Higgs superfield
with the same gauge quantum number  results in the first two
terms of
the following $R$-parity-violating superpotential, while the third
term is also not forbidden by any symmetry,
\begin{equation}
{\cal W}_{\not R} =  \lambda_{ijk} L_i L_j E^c_k
                      +  \lambda'_{ijk} L_i Q_j D^c_k
                      +  \lambda''_{ijk} U^c_i D^c_j D^c_k  ,
\label{R-parity}
\end{equation}
where $L_i$ and $Q_i $ are the $SU(2)$-doublet lepton and quark
superfields and $E^c_i, U^c_i, D^c_i$ are the singlet  superfields;
$\lambda_{ijk}$ is antisymmetric under the interchange of the first
two indices, while $\lambda''_{ijk}$ is antisymmetric under the
interchange of the last two. It is obvious that there are 45 such
Yukawa couplings:~9 each of the $\lambda$ and $\lambda''$ types and
27 of the $\lambda'$ variety. These couplings cannot all be arbitrary
simultaneously and have to be {\it carefully} instigated so that the
consistency with various experimental measurements is respected
\cite{rpar}. This requires that some of them are not dynamically
active
at the same time. For example, assuming all the $\lambda''$ couplings
to be zero, which we adopt in any case for the rest of this paper,
the
bounds from the non-observation of proton decay and $n$--$\bar{n}$
oscillation are avoided. This assumption also makes it simpler to
evade the cosmological bounds \cite{cosm}.
The $L$-violating couplings can, in
principle, wash out the GUT-scale baryogenesis, but Dreiner and Ross
\cite{ross} have argued that these bounds are highly model-dependent.
If one of the $L$-violating couplings involving a particular lepton
family is small ($< 10^{-7}$), so as to conserve the corresponding
lepton flavour over cosmological time scales, then the primordial
baryogenesis could be restored and, therefore, these bounds would no
longer be effective.

\vskip 10pt

The essential idea of this paper is the following:~ In the SM,
one of the semileptonic decay channels of the $b$-quark
is $b\rightarrow ce\bar{\nu}$. In the MSSM, if one relaxes the
assumption of $R$-parity
conservation minimally, assuming only one of the above-mentioned 45
Yukawa couplings, namely, $\lam$, to be non-zero\footnote{ This
tacit assumption that only one $R$-parity-violating coupling is
non-zero
at a time -- a standard practice in ${\not \! \!{R}}$-phenomenology
-- is in accordance with a non-GUT set up, as otherwise if
${\not \! \!{R}}$ is embedded in a GUT scenario, one cannot
avoid various $R$-parity-violating couplings occuring simultaneously
from the GUT-multiplet structures.},
and if the photino lies in the mass range ($\sim$ 2--3 GeV),
then the $b$-quark can have a
$b\rightarrow ce\phot$ decay mode. For the sake of simplicity we
consider the photino ($\phot$) as the lightest neutralino, which we
argue later to be stable within the size of the relevant detectors.
The $\phot$ will mimic the behaviour of the SM $\nu$ providing an
invisible channel of energy;
however, owing to its massive nature it will
arrange a different  kinematical configuration of the visible decay
products compared to the SM scenario.
The magnitude of $\lam$ that we
allow for a given value of the photino mass ($m_{\phot}$) is of
course constrained from the experimental measurement of the
semileptonic branching ratio. Our search strategy lies in
comparing and contrasting the differential distributions in the two
cases which, due to the difference in their kinematic configurations,
could unveil the new physics signal to an observable scale.

\vskip 10pt

We start with the consideration of $B$-meson decays at the
quark level. In the SM, the decay matrix element of
$b(P)\rightarrow c(k_2)+e(k_1)+\bar{\nu_e}(k_3)$ and its
spin-sum-square are given by
\begin{eqnarray}
{\cal{M}} &=& 2\sqrt{2} G_F~ V_{cb}~ \bar{e}(k_1)~{\gamma_\mu}
            ~P_L~\nu_e(-k_3)~\bar{c}(k_2)~\gamma^\mu~P_L~b(P),
              \nonumber \\
\Sigma_{spin}|{\cal{M}}|^2 &=& 128~G_F^2~
|V_{cb}|^2~ (k_1\cdot k_2)(k_3\cdot P),
\end{eqnarray}
where the symbols have their usual significance.  We neglect the
electron mass and introduce $r_c = {m_c}/{m_b}, ~
x = {2E_e}/{m_b}$, where $E_e$ is the energy of the electron
in the $b$-quark rest frame. We obtain,
\begin{eqnarray}
\frac{d\Gamma^{\rm SM}}{dx} & = & \frac{G_F^2 m_b^5}{192\pi^3}
|V_{cb}|^2~f(x,r_c), \nonumber \\
\Gamma^{\rm SM} & = & \int_0^{1-r_c^2} dx
{}~\frac{d\Gamma^{\rm SM}}{dx},
\end{eqnarray}
where
\begin{equation}
f(x,r_c) = \frac{2x^2}{(1-x)^3} (1-x-r_c^2)^2 \left(3-5x+2x^2
                               -r_c^2x+3r_c^2\right).
\end{equation}

\vskip 10pt

Now we turn our attention to the $R$-parity breaking sector. The
part of the Lagrangian induced by $\lam$, the only one relevant
for our purpose, is given by
\begin{equation}
{\cal{L}} = \lam\left[\bar{b}_R \nu_{eL} \tilde{s}_L
+\bar{b}_R s_L \tilde{\nu}_{eL}
-\bar{b}_R e_L \tilde{c}_L -\bar{b}_R c_L \tilde{e}_L
+\overline{(\nu_{eL})^c} s_L \tilde{b}_R^*
-\overline{(e_L)^c} c_L \tilde{b}_R^* \right] + {\rm h.c.}
\end{equation}
The new process, which mimics the semileptonic $b$-decay in the
present situation, is
\begin{equation}
b(P) \rightarrow c(k_2) + e(k_1) + \phot(k_3).
\end{equation}
The above interaction proceeds through two different
interfering channels.
In one case, the $b$-quark decays to a $c$-quark
and a virtual $\sel$ induced by $\lam$ and the
latter in turn decays with
an electromagnetic strength to a $\phot$ and an electron.
The decay matrix element is given by
\begin{equation}
{{\cal{M}}_{1\not{R}}} = \frac{\sqrt{2}e\lam}{m_{\sel}^2}
\left[\bar{e}(k_1)~P_R~\phot(-k_3)~~\bar{c}(k_2)~P_R~b(P)\right].
\end{equation}
In the other case, the $b$-quark emits a $\phot$ and a virtual
$\bsq$ and the latter decays into a $c$-quark and an electron
via $\lam$. The decay matrix element in this case is
given by\footnote{The sign difference between
${{\cal{M}}_{1\not{R}}}$ and ${{\cal{M}}_{2\not{R}}}$ has its
root in
the sign difference between the photino couplings to the left-
and the right-type scalars \cite{haberkane}.}
\begin{equation}
{{\cal{M}}_{2\not{R}}} = -\frac{\sqrt{2}e\lam}{m_{\bsq}^2}
\left[\bar{c}(k_2)~P_R~e^c(-k_1)~~\bar{\phot}(k_3)~P_R~b(P)\right].
\end{equation}
Using a simple Fierz transformation, we obtain the spin-summed
and squared matrix element as
\begin{eqnarray}
\Sigma_{spin} |{\cal{M}}_{1\not{R}} + {\cal{M}}_{2\not{R}}|^2 &=&
\left(\frac{2\sqrt{2}e\lam}{m_{\sel}^2}\right)^2
 \left[(P\cdot k_2)(k_1\cdot k_3)
+ \kappa^4 (P\cdot k_3)(k_1\cdot k_2) \right.
\nonumber \\
  &\! \! \! \! \! \! - & \! \! \! \! \! \! \! \! \left. \kappa^2
\left\{(P\cdot k_2)(k_1\cdot k_3) + (P\cdot k_3)(k_1\cdot k_2) -
     (P\cdot k_1)(k_2\cdot k_3)\right\}\right],
\end{eqnarray}
where $\kappa = m_{\sel}/m_{\bsq}$. Introducing $r_{\phot}
= m_{\phot}/{m_b}$, the contribution to the $R$-parity
induced width can be written as
\begin{eqnarray}
\frac{d\Gamma_{\not{R}}}{dx} &=& \frac{G_F^2 m_b^5}{192\pi^3}
{}~\left(\frac{e\lam}{g^2}\right)^2~\left(\frac{m_W}{m_{\sel}}\right)^4
 ~8~ \lambda^{1/2}\left(1,\frac{r_c^2}{1-x},\frac{r_{\phot}^2}{1-x}
\right) ~x^2
 \left[\left\{ f_1(1+\kappa^2)^2
\right.\right. \nonumber \\
& & \left.\left.
- f_2 \kappa^2\right\}(1-x) - f_2 (1-\kappa^2)^2
\left(1-\frac{x}{2}\right)\right],  \\
 \Gamma_{\not{R}} &=& \int_0^{1-(r_c+r_{\phot})^2}~dx~
\frac{d\Gamma_{\not{R}}}{dx},
\end{eqnarray}
where
\begin{eqnarray}
\lambda(a,b,c) & = & a^2 + b^2 +c^2 - 2ab - 2bc -2ca,\\
f_1  & = & \frac{1}{2} - \frac{r_c^2+r_{\phot}^2}{1-x}
 + \frac{1}{2} \frac{(r_c^2-r_{\phot}^2)^2}{(1-x)^2}, \\
f_2  & = &
-1 - \frac{r_c^2+r_{\phot}^2}{1-x}
+ 2 \frac{(r_c^2-r_{\phot}^2)^2}{(1-x)^2}.
\end{eqnarray}

Now we discuss the results of our calculation:
\begin{enumerate}
\item
We calculate the total (SM $+ {\not \! \! R}$) semileptonic branching
ratio $B(b\rightarrow c+e+{\rm invisible})$ and allow only as much
admixture of $\lam$ as saturates the uncertainty of the corresponding
global average $(10.43\pm0.24)\%$ \cite{pdg}\footnote{ We stay on
the conservative side by using the average which assumes leptonic
universality. We also note that the new graphs we consider
contribute to the semi`electronic' $b$-decay mode only, thus
violating the leptonic universality. However, the experimental
uncertainty of each of the lepton flavour specific channels
(see ref. \cite{pdg} for details) is larger than the average and,
therefore, the above universality violation is easily tolerable.}.
In the absence of a precise SM
prediction of the above, this approximation is reasonable.
We display in Fig. 1 the $90\%$ C.L. upper bound on the $\lam$ as a
function of $m_{\phot}$ for fixed $m_{\bsq} = 100$ GeV and for three
different values of $m_{\sel} = 50, 100$ and 200 GeV, corresponding
to $\kappa = 0.5, 1$ and 2, respectively.
It may be noted
that the existing $1\sigma$ limit $\lam <0.26$  from the
forward--backward asymmetry measurements in $e^+e^-$ collisions,
derived in \cite{bgh}, is based on an effective operator which goes
like ${\lambda'}/{\widetilde{m}}$, whereas in our case it scales like
${\lambda'}/{\widetilde{m}}^2$, $\widetilde{m}$ being a common scalar
mass parameter.
We also note that the Cabibbo-Kobayashi-Maskawa element $V_{cb}$
enters the SM part of the calculation which is otherwise extracted
from the semileptonic measurements \cite{kim}.
Any non-standard contamination in the
process would certainly affect the extraction.
We keep our inclusive branching ratio consistent with
experiment by assuming $V_{cb} = 0.04$ and saturating the uncertainty
with $\lam$ as mentioned earlier. We keep $m_b = 5$ GeV in our
calculation.

\item
For given values of photino- and scalar-masses, imposing the
$90\%$ C.L. limit of $\lam$ from Fig. 1, we exhibit in Fig. 2 the
electron energy distribution, {\it i.e.} ${\frac{1}{\Gamma}}
{\frac{d\Gamma}{dx}}$, where $\Gamma$ corresponds to $(i)$
$\Gamma_{\rm SM}$ and $(ii)$ $\Gamma_{\rm tot} = \Gamma_{\rm SM}
+ \Gamma_{\not {R}}$.
We notice that for fixed $m_{\phot}$ and
$m_{\bsq}$, varying $m_{\sel}$ does not show any observable impact
within the scale of the figures. More important in determining the
shapes of the distributions are the external masses which affect
the kinematical configurations and, of course, the scale of the
new physics given by either $m_{\bsq}$ or $m_{\sel}$; much less
important in this case is the relative size of $m_{\bsq}$ and
$m_{\sel}$, which affects the dynamics of the new graphs.

\item
In Fig. 3, we demonstrate the angular distribution (angle between
the $c$-quark and the electron) for the two cases stated above.
Here, too, we find that the shapes of the distributions are quite
insensitive to the relative scales of $m_{\bsq}$ and $m_{\sel}$,
as in Fig. 2.

\end{enumerate}

Two important issues need to be addressed in any discussion
concerning semileptonic $B$-decays. One is the QCD correction
and the other the hadronization effect. In the SM, the
QCD  corrections have been computed for massless
leptons \cite{qcd1} and have been recently estimated \cite{haber}
in the context of the 2-Higgs doublet model.
The ${\cal{O}}(\alpha_S)$
correction reduces the SM prediction by $\sim 10\%$. In our case,
given the smallness of $\lam$,
the QCD corrections to the new graphs are unlikely to make a
significant impact on our estimate. One also notes that the
$b$-quark hadronizes much before it decays weakly. However, in a
heavy quark system the spectator model works in a very reasonable
way, which is evident from the consistency between the lifetime
measurements of various $B$-hadrons \cite{pdgb}.
At LEP, the $b$-quarks are energetic (45 GeV).
But the distributions in the $b$-quark rest frame shown in our
analysis can be readily compared with the observations made in the
laboratory. The high-resolution
silicon microvertex detector makes it possible to locate the
primary decay vertex of the $B$-hadron from the direction of the
$c$-quark jet and the charged lepton. Once the decay vertex is
obtained, the flight direction of the $b$-quark is known and since
its energy is fixed, the relevant boost factors
relating the $b$-quark rest frame and the laboratory frame
can be calculated.

\vskip 10pt

At this stage we argue in favour of two essential hypotheses
that we advocated in the beginning. We assumed that (i) the photino
lifetime is not more than 1 s, meaning $\Gamma_{\phot} \geq
10^{-22}$ MeV and (ii) the photino is invisible, {\it i.e.} it does
not decay within a few metres, implying
$\Gamma_{\phot} \leq 10^{-13}$
MeV in its rest frame\footnote{To anticipate the latter number,
assume, as an example, that a $\phot$ is produced at rest in the
$b$ rest frame and consider a $b$-quark of mass 5 GeV in a 45 GeV
jet at LEP. Now,
counting a boost factor of 9, the $\phot$ must have a life-time of
$10^{-9}$ s in its rest frame, {\it i.e.} a width of $6.6 \times
10^{-13}$ MeV, to travel at least $\sim 3$ m in the lab frame to
escape detection.}. Now with $\lam$ as the only $R$-parity breaking
coupling, there are essentially two obvious decay modes of the
photino. In the first case, the decay of the photino would go via
the flavour-violating
photino-quark-squark coupling\footnote {It arises from the mismatch
between the squark and quark mass-matrices and is strongly restricted
by flavour-changing neutral current constraints.},
where at one vertex the
$\phot$ decays to an $s$-antiquark and a virtual $b$-squark while the
latter at the other vertex decays preferentially to an $s$-quark
and a $\nu$ {\it via} $\lam$.
We define the flavour-violating parameter
`$c$' by scaling `$e$' to `$ce$' in the photino-quark-squark vertex.
Then, for a photino of mass 2 (3) GeV, assuming $m_{\bsq} = 100$
GeV, $\kappa
= 1$ and reading the corresponding $\lam$ values from Fig. 1, we
estimate $\Gamma_{\phot} \simeq 3c^2 \times 10^{-13}~ (1.3c^2 \times
10^{-10})$ MeV. Suffice it to choose, therefore, $c \leq 0.03$ to
enable the photino to fly safely out of the detector\footnote{ It
should be remembered that in a detector, CLEO for example, where
the $B$'s are produced almost at rest, the boost factor of 9
counted for LEP is absent. Hence, the constraint on $c$ should
actually be a factor of $\sim 3$  stronger than quoted in
the text.}.
Indeed, the
above estimates are somewhat crude and might be modified by
hadronization effects, but are sufficient
to demonstrate that a $\phot$ can
very well act  as an invisible particle even in a
$R$-parity-violating atmosphere.  The other decay channel of the
$\phot$ is characterized by
its splitting to a virtual $b$-quark and a virtual $b$-squark and
their subsequent decays (the latter {\it via} $\lam$). These two
different types of decays proceed with roughly similar strength.
We note that with the above choice of `$c$', the radiative decay
of the photino ($\phot \rightarrow \gamma \nu_e$ penguin) is well
under control.  We have also
checked that the parameters can be adjusted reasonably well to ensure
the other limit, {\it i.e.} $\Gamma_{\phot} \geq10^{-22}$ MeV.

\vskip 10pt

Since in our scenario the LSP is stable inside a realistic detector
even though $R$-parity is violated, the canonical LSP search
strategy should, in principle, apply to our case as well.
For example, the
L3 Collaboration at LEP in a recent analysis \cite{l3} have
excluded an LSP weighing below 18 GeV. However, this bound
evaporates if $\tan\beta < 2$; moreover, the above analysis
relies on the GUT-relation between the gaugino masses.
The OPAL Collaboration at LEP \cite{opal}
have looked for massive photinos
decaying very fast within the detector
{\it via} a $\lambda_{123}$-type coupling and excluded
$m_{\phot} =$4--43 GeV for $m_{\sel} < 42$ GeV,
and $m_{\phot} =$ 7--30 GeV for $m_{\sel} < 100$ GeV
(95$\%$ C.L.). For such a fast decaying photino,
they could not look for the window ($m_{\phot}<4$ GeV), since the
$\phot$-pairs cannot be separated from the $\tau$-pairs. The
ALEPH Collaboration at LEP \cite{aleph}, dealing with a more
general $\lambda$-type coupling and considering a general LSP
rather than a pure photino, have improved the above exclusion zone
and have also reported their negative results on other
supersymmetric particles up to their kinematic limit ($<m_Z/2$). Our
proposed mode of searching for a photino lighter than 4 GeV with a
$\lam$-type coupling,
therefore, covers a complementary zone in the supersymmetric
parameter
space.  We point out though that our analysis relies on a simple
assumption that the LSP is dominantly a photino.

\vskip 10pt

We now comment how our analysis could be
extended for a few other $\lambda'$-type couplings as well.
Considering the fact that a $\mu$ in the final state in a
semileptonic $B$-decay constitutes as viable a mode for detection
as an $e$ in the final state, taking care of the slight
modification in the kinematics due to the $\mu$-mass, our analysis
could be carried out also for $\lambda'_{223}$, whose existing
bound is 0.16 ($D$-decay, 2$\sigma$) \cite{bc}.
$\lambda'_{323}$
should be handled somewhat differently for phase-space consideration
and also because the $\tau$ decaying within the detector would
change the signal profile. Consideration of $\lambda'_{113}$ in
our analysis boils down to a replacement of a $c$-quark jet with
a $u$-quark jet (the latter cannot be distinguished from a general
hadronic activity): $V_{ub}$ being  much smaller than $V_{cb}$
provides less SM background than the situation involving
$\lam$. Similarly, $\lambda'_{213}$ could also be utilised for a
$\mu$ in the final state.
The existing limits \cite{bgh} on $\lambda'_{113}$ and
$\lambda'_{213}$ are 0.03 (charged-current universality, 2$\sigma$)
and 0.09 ($\pi$-decay, 1$\sigma$) respectively. In a very recent
analysis \cite{ag}, $K^+\rightarrow \pi^+ \nu \bar{\nu}$ has been
used to place stringent constraints on all of $\lam$,
$\lambda'_{223}$, $\lambda'_{113}$ and $\lambda'_{213}$. The bound
on each of them is $0.012 (m_{{\tilde d}_{Rk}}/100~{\rm GeV})$
at 90$\%$ C.L. As emphasized earlier,  the crucial feature of our
analysis lies in the kinematic properties.
Moreover, each of the $\sel$-exchanged and $\bsq$-exchanged processes
of our analysis contributes with roughly the same magnitude, so that
making $\bsq$ heavy -- thereby evading the $K^+$-decay bound -- while
keeping $\sel \sim 100$ GeV, we have checked does not change our
results much.

\vskip 10pt

Finally, we comment that the `semileptonic anomaly', namely the
long-standing {\it irritation}
that the SM prediction of the semileptonic branching ratio lies
somewhat above what has experimentally been observed, still exists
of course amidst
various theoretical as well as experimental uncertainties.
We admit that the picture considered above instead of curing the
anomaly worsens it further, since the new process adds
incoherently to the SM graph.
Nevertheless, we keep ourselves strictly
consistent by admitting only that much $\lam$ which does not let
the prediction of the inclusive branching ratio exceed its
$90\%$ C.L. observation.

\vskip 10pt

The kind of scenario that
we have dealt with in this paper, arguably lies in a corner of
the vast supersymmetric parameter space, yet has the virtue of
manifesting itself through a simple study of the kinematic
configurations of the semileptonic decay products.
It cannot be denied that $R$-parity violation and the simultaneous
presence of a light photino might appear as a somewhat contrived
scenario, although we have tried to motivate the former from the
latter.
But this could very well turn out to be a reality and one must ensure
that it does not slip through the canonical supersymmetry search
biased by a thick layer of theoretical prejudice of $R$-parity
conservation and Grand Unification.  Interestingly, the scenario,
as we
have demonstrated in this work, is very much within the $B$-physics
reach at LEP or at CLEO. A thorough study incorporating the
hadronization effect and implementing the full detector simulation
appropriate to these colliders is, therefore, called for.

\vskip 10pt

We thank T. Aziz, S. Banerjee, D. Choudhury, S. Dutta, G. Giudice
and M. Maity for fruitful discussions. AR is grateful to the Theory
Division of CERN for hospitality when the work was initiated. His
research is supported by the Council of Scientific and Industrial
Research, India, and the Department of Science and Technology, India.

\newpage

\setcounter{figure}{0}
\begin{figure}[htbp]
\vskip 9.0in\relax\noindent\hskip -1in\relax
{\includegraphics{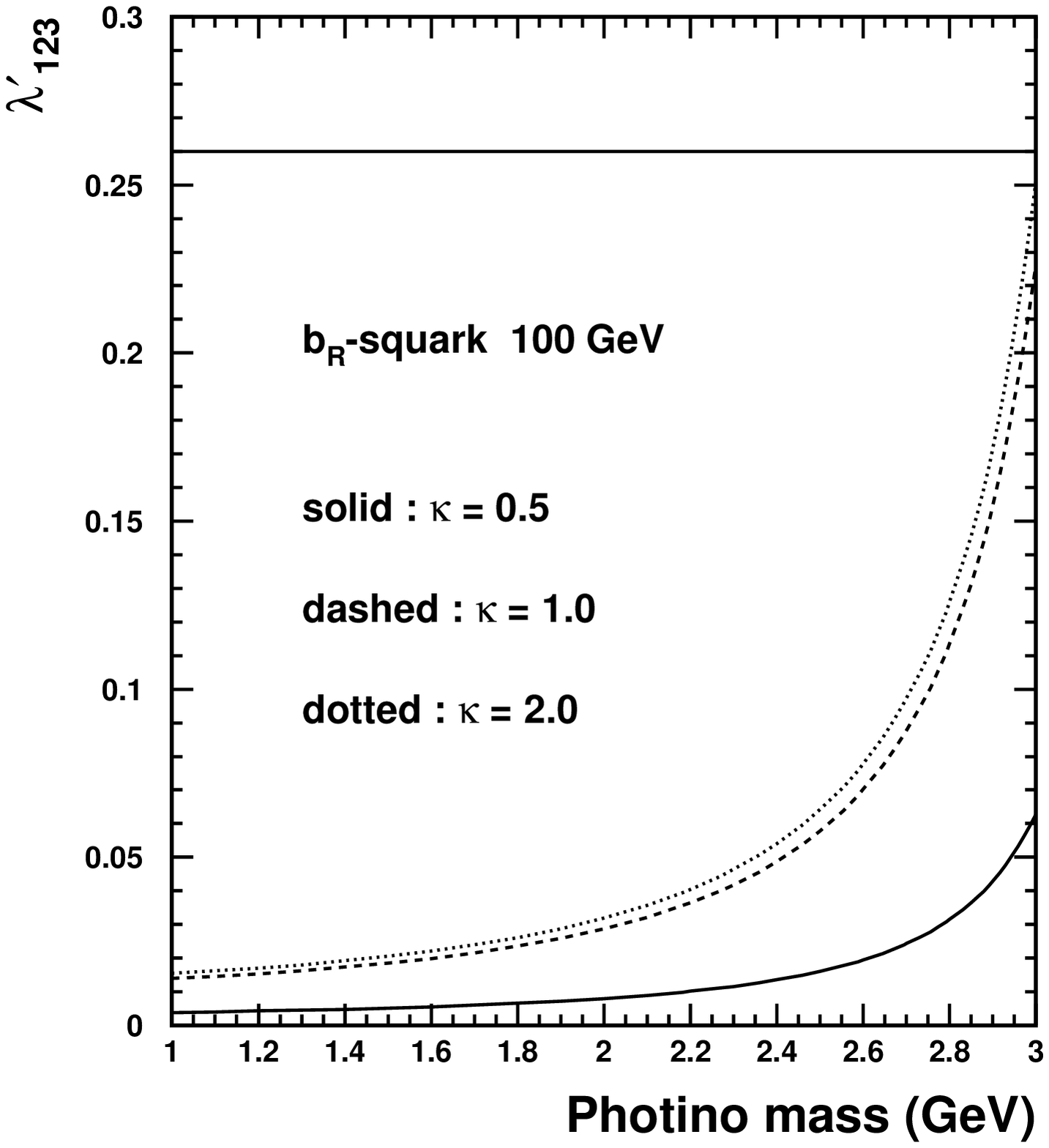}}
\vskip -2in\caption{The 90$\%$ upper limit on $\lam$ as a function of
the photino mass for various combinations of scalar masses. For fixed
$m_{\bsq}$ = 100 GeV, $\kappa = 0.5,~1,~2$ correspond to $m_{\sel} =
50,~100,~200$ GeV, respectively.  The horizontal line is
the $1\sigma$ bound derived in ref. [8] from forward--backward
asymmetry of $e^+e^-$ collisions at low energy.}
\end{figure}

\newpage

\begin{figure}[htbp]
\vskip 9.0in\relax\noindent\hskip -1in\relax
{\includegraphics{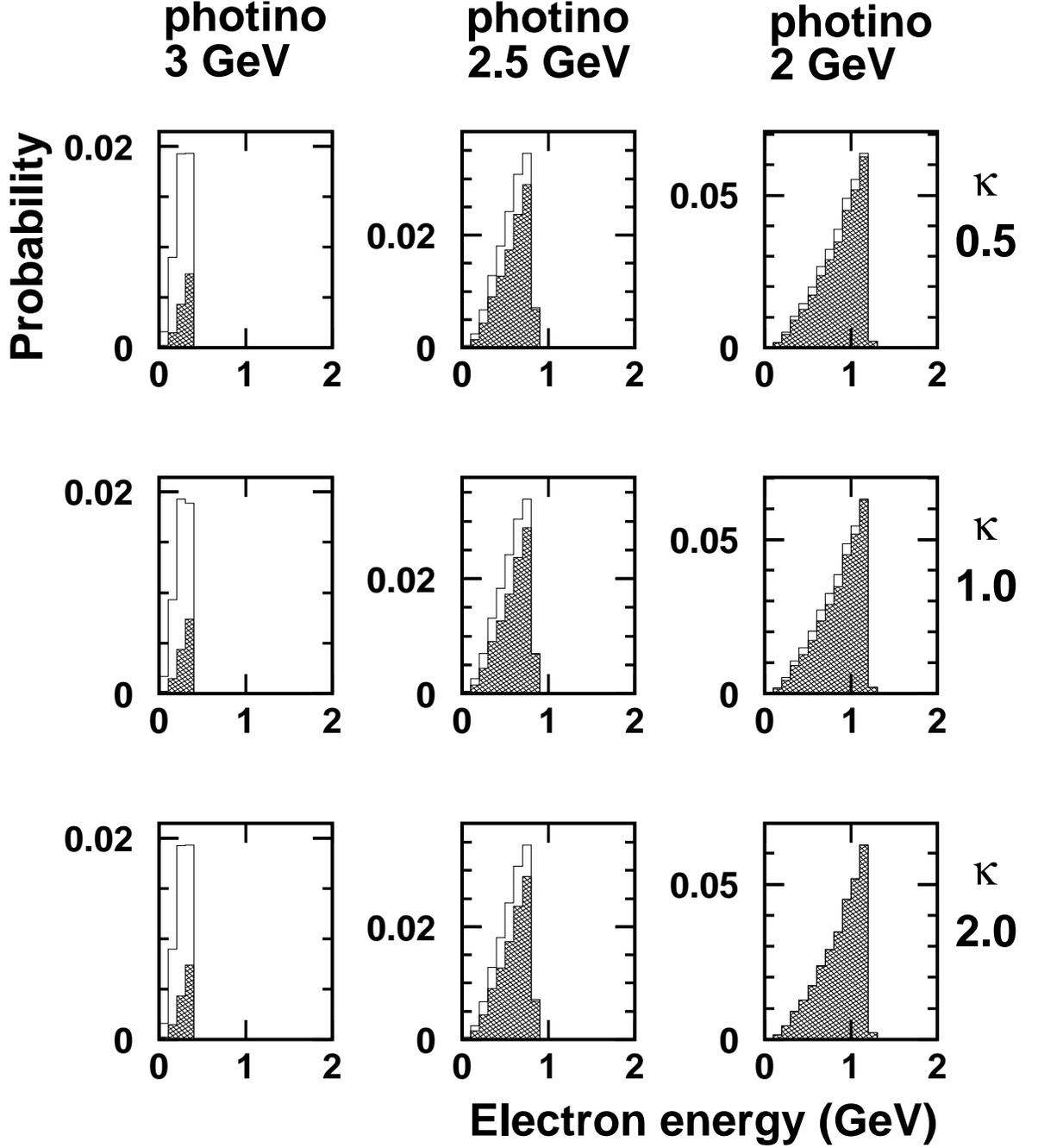}}
\vskip -2in\caption{The electron-energy distribution
in nine different cases.
Those SM events which are within the kinematic boundary of the
$R$-parity-violating mode are shown by the shaded area.
In a given column these figures correspond to
various choices of $\kappa$ (as in Fig. 1)
for a fixed $m_{\phot}$, while
in a given row they correspond to different values of
$m_{\phot}$ for a fixed $\kappa$.}
\end{figure}

\newpage

\begin{figure}[htbp]
\vskip 9.0in\relax\noindent\hskip -1in\relax
{\includegraphics{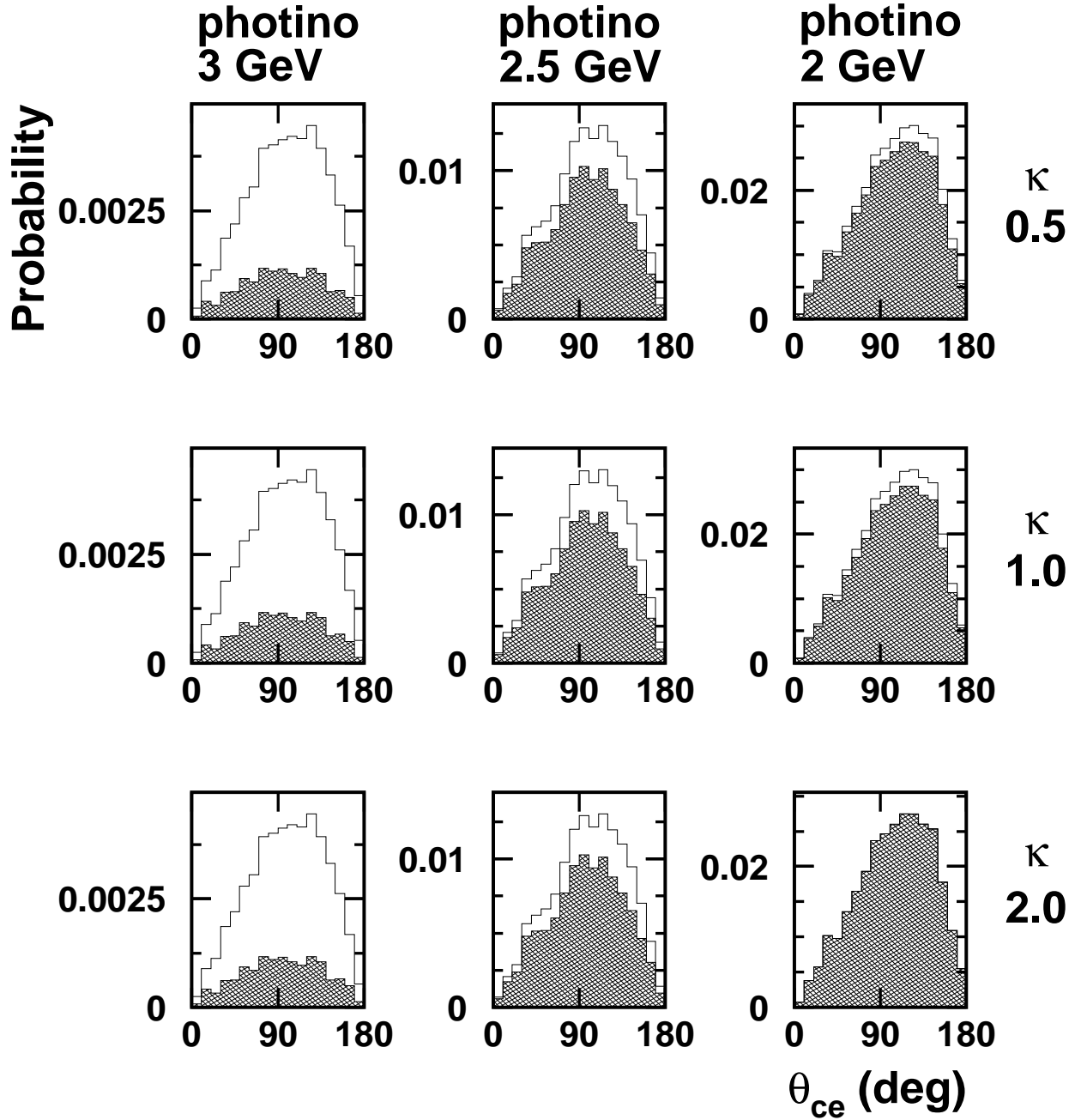}}
\vskip -2in\caption{The charm quark--electron angle distribution
in nine different cases. As in Fig. 2, SM events corresponding to
electron energies  within the allowed kinematic boundary of the
$R$-parity-violating mode are shown by the shaded area.
In a given column these figures correspond to
various choices of $\kappa$ (as in Fig. 1) for a fixed $m_{\phot}$,
while in a given row they correspond to different values of
$m_{\phot}$ for a fixed $\kappa$.}
\end{figure}

\end{document}